\documentclass{emulateapj}

\newcommand{\myemail}{michal@dark-cosmology.dk}

\slugcomment{Accepted in ApJ on Nov 13 2008}

\shorttitle{The host galaxy of GRB 980425 / SN 1998bw}
\shortauthors{Micha{\l}owski et al.}

\begin{document}

\title{The properties of the host galaxy and the immediate environment of GRB 980425 / SN 1998bw from the multi-wavelength spectral energy distribution}

\author{Micha{\l}~J.~Micha{\l}owski\altaffilmark{1}, 
	Jens~Hjorth\altaffilmark{1}, 
	Daniele~Malesani\altaffilmark{1},
	Tadeusz~Micha{\l}owski\altaffilmark{2},
	Jos\'{e}~Mar\'{i}a~Castro~Cer\'{o}n\altaffilmark{1},
	Robert~F.~Reinfrank\altaffilmark{3,4},
	Michael~A.~Garrett\altaffilmark{5,6,7},
	 Johan~P.~U.~Fynbo\altaffilmark{1},
	Darach~J.~Watson\altaffilmark{1}
and 
	Uffe~G.~J\o rgensen\altaffilmark{8}
	}

\altaffiltext{1}{Dark Cosmology Centre, Niels Bohr Institute, University of Copenhagen, Juliane Maries Vej 30, DK-2100 Copenhagen \O, Denmark}
\altaffiltext{2}{Astronomical Observatory, Adam Mickiewicz University, S\l oneczna 36, 60-286 Pozna\'n, Poland}
\altaffiltext{3}{Australia Telescope National Facility, CSIRO, P.O. Box 76, Epping, NSW 1710, Australia}
\altaffiltext{4}{School of Chemistry \& Physics, The University of Adelaide, Adelaide, SA 5005, Australia}
\altaffiltext{5}{Netherlands Institute for Radio Astronomy (ASTRON), Postbus 2, 7990 AA Dwingeloo, The Netherlands}
\altaffiltext{6}{Leiden Observatory, University of Leiden, P.B. 9513, Leiden 2300 RA, the Netherlands}
\altaffiltext{7}{Centre for Astrophysics and Supercomputing, Swinburne University of Technology, Hawthorn, Victoria 3122, Australia}
\altaffiltext{8}{Niels Bohr Institute, University of Copenhagen, Juliane Maries Vej 30, DK-2100 Copenhagen \O, Denmark}
\email{\myemail}

\begin{abstract}
We present an analysis of the spectral energy distribution (SED) of the galaxy ESO 184-G82, the host of  the closest known long gamma-ray burst (GRB) 980425 and its associated supernova (SN) 1998bw. We use our observations obtained at the Australia Telescope Compact Array (the third $>3\sigma$ radio detection of a GRB host) as well as archival infrared  and ultraviolet (UV) observations to estimate its star formation state.  We find that  ESO 184-G82 has  a UV star formation rate (SFR) and stellar mass consistent with the population of cosmological GRB hosts and of local dwarf galaxies. However, it has a higher specific SFR (per unit stellar mass) 
than luminous spiral galaxies. The mass of ESO 184-G82 is dominated by an older stellar population in contrast to the majority of GRB hosts. The Wolf-Rayet region $\sim800$ pc from the SN site experienced a starburst episode during which the majority of its stellar population was built up. Unlike that of the entire galaxy, its SED is similar to those of cosmological submillimeter/radio-bright GRB hosts with hot dust content.  These findings add to the picture that in general, the environments of  GRBs on 1--3 kpc scales are associated with high specific SFR and hot dust. 
\end{abstract}


\section{Introduction}

Long gamma-ray bursts (GRBs) are associated with the death of  massive stars \citep[e.g.][]{galamanature,hjorthnature,stanek}. This makes them of special interest in cosmology because they possibly trace the evolution of the rate of star formation in the Universe \citep[e.g.][]{lambreichart00,jakobsson05,jakobsson06}. Indirect evidence of the nature of GRBs was found by studying their host galaxies  \citep[e.g.][]{bloom,christensen04,sollerman05,castroceron08,savaglio09}.  

Moreover, several studies on the immediate environments of GRBs suggest a close connection of long GRBs with regions of star formation, and therefore that their progenitors are likely massive stars. \citet{fruchter06} found that GRBs trace the ultraviolet (UV) brightest parts of their host \citep[see also][]{bloom02}. 
\citet{thone08} studied in detail the environment (in $3$ kpc bins) of \object{GRB 060505}, concluding that it originated in the youngest, most metal-poor and most intense star-forming region in the host galaxy. Similarly, \citet{ostlin08} found that the $0.3$ kpc environment of \object{GRB 030329} is much younger than the entire galaxy and its estimated age suggests a conservative lower limit on the mass of the GRB progenitor equal to $12\,M_\odot$. Finally, a significant number of other GRBs were reported to reside in dense star-forming regions \citep{castrotirado99,hollandhjorth99,hjorth03,savaglio03,vreeswijk04,chen05,chen06,fynbo06b,watson06,watson07, prochaska07b,prochaska07,ruizvelasco07} and molecular clouds \citep{galamawijers01,stratta04,jakobsson06b,campana07,prochaska08}.

\object{GRB 980425} is the closest known GRB \citep[$z=0.0085$;][]{tinney98}, therefore it is an excellent laboratory for local GRB studies.
\citet{galamanature} reported the Type Ic supernova \object[SN 1998bw]{(SN) 1998bw} exploding inside the error box of \object{GRB 980425}. Its lightcurve was well modeled by an explosion of a Wolf-Rayet (WR) star \citep{iwamoto98}, which is a highly evolved and massive star that has lost its outer hydrogen layers. Up to now, three other GRBs have also been spectroscopically confirmed to be associated with SNe: \object{GRB 030329} \citep{hjorthnature,matheson03,stanek}, \object[GRB 031203]{031203} \citep{cobb04,galyam04,malesani04,thomsen04} and \object[GRB 060218]{060218} \citep{ferrero06,mirabal06,modjaz06,pian06,soderberg06nature,sollerman06},  while two GRBs were confirmed to be SN-less: GRB \object[GRB 060505]{060505} and \object[GRB 060614]{060614} \citep{fynbo06,dellavalle06,galyam06}. 

The host galaxy of \object{GRB 980425} / \object{SN 1998bw}  \citep[\object{ESO 184-G82};][]{holmberg77} is a dwarf \citep[$0.02$ of the characteristic blue luminosity, $L_B^*$;][]{fynbo00} barred spiral  \citep[SBc;][]{fynbo00} with axis diameters of $12$ and $10$ kpc \citep[down to $B=26.5$ mag arcsec$^{-2}$;][]{sollerman05}, dominated by a large number of star-forming regions \citep{fynbo00,sollerman05}. 
\object{SN 1998bw} occurred inside one of these, $\sim800$ pc southeast of a region displaying a Wolf-Rayet type signature spectrum  \citep[hereafter: WR region;][]{hammer06}. The WR region dominates the galaxy's emission at $24\,\micron$  \citep{lefloch} and is the youngest region within the host  exhibiting very low metallicity  \citep{christensen08}.

In this paper we present fits to the spectral energy distribution (SED) of \object{ESO 184-G82} and the WR region and compare their properties to other galaxies. Section \ref{sec:data} lists the data sources \citep[including the third radio detection of a GRB host after those reported by][]{bergerkulkarni,berger} used for the SED modeling of Section \ref{sec:model}. We derive properties of the host galaxy and WR region in Section \ref{sec:results}, discussing their implications in Section \ref{sec:discussion}. Section \ref{sec:conclusion} closes with our conclusions.
We use a cosmological model with $H_0=70$ km s$^{-1}$ Mpc$^{-1}$, 
$\Omega_\Lambda=0.7$ and $\Omega_m=0.3$, so \object{ESO 184-G82} is at a luminosity distance of  $36.5$ Mpc and $1''$ corresponds to 175 pc at its redshift.

\section{Data}
\label{sec:data}

\begin{figure}
\begin{center}
\plotone{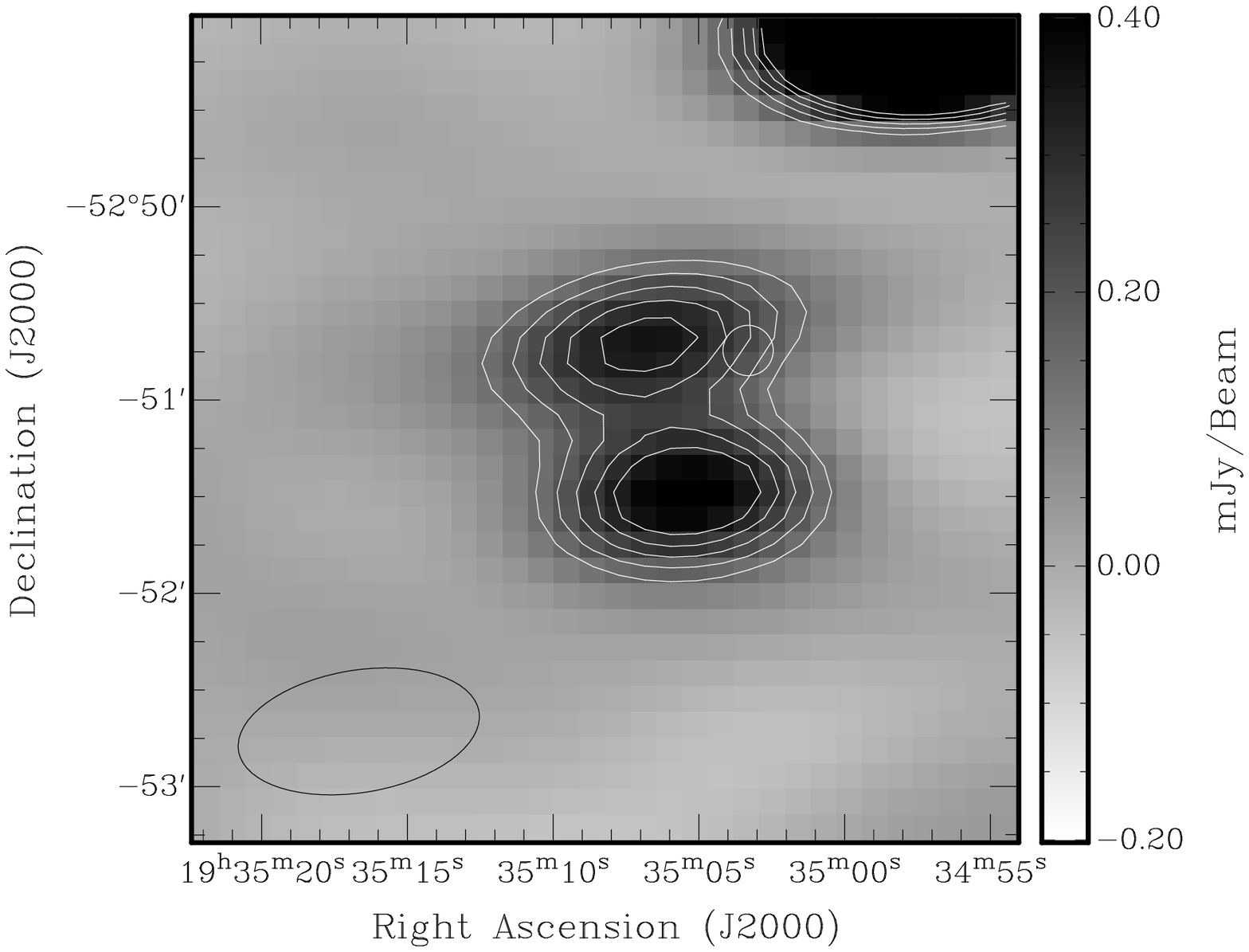}
\end{center}
\caption{6 cm ATCA image with size of $5.7'$ or $60$ kpc at the redshift of $0.0085$. The circle marks the position of \protect\object{SN 1998bw}. The two overlapping objects in the middle are the  \protect\object{ESO 184-G82} (north) and galaxy A of \citet{foley06} (south).  The contours are $3$, $4$, $5$, $6$, and $7\sigma$, where $\sigma=46\,\mu$Jy beam$^{-1}$. The beam is shown in the bottom left corner.}
\label{fig:radio}
\end{figure}

\begin{figure*}
\begin{center}
\plotone{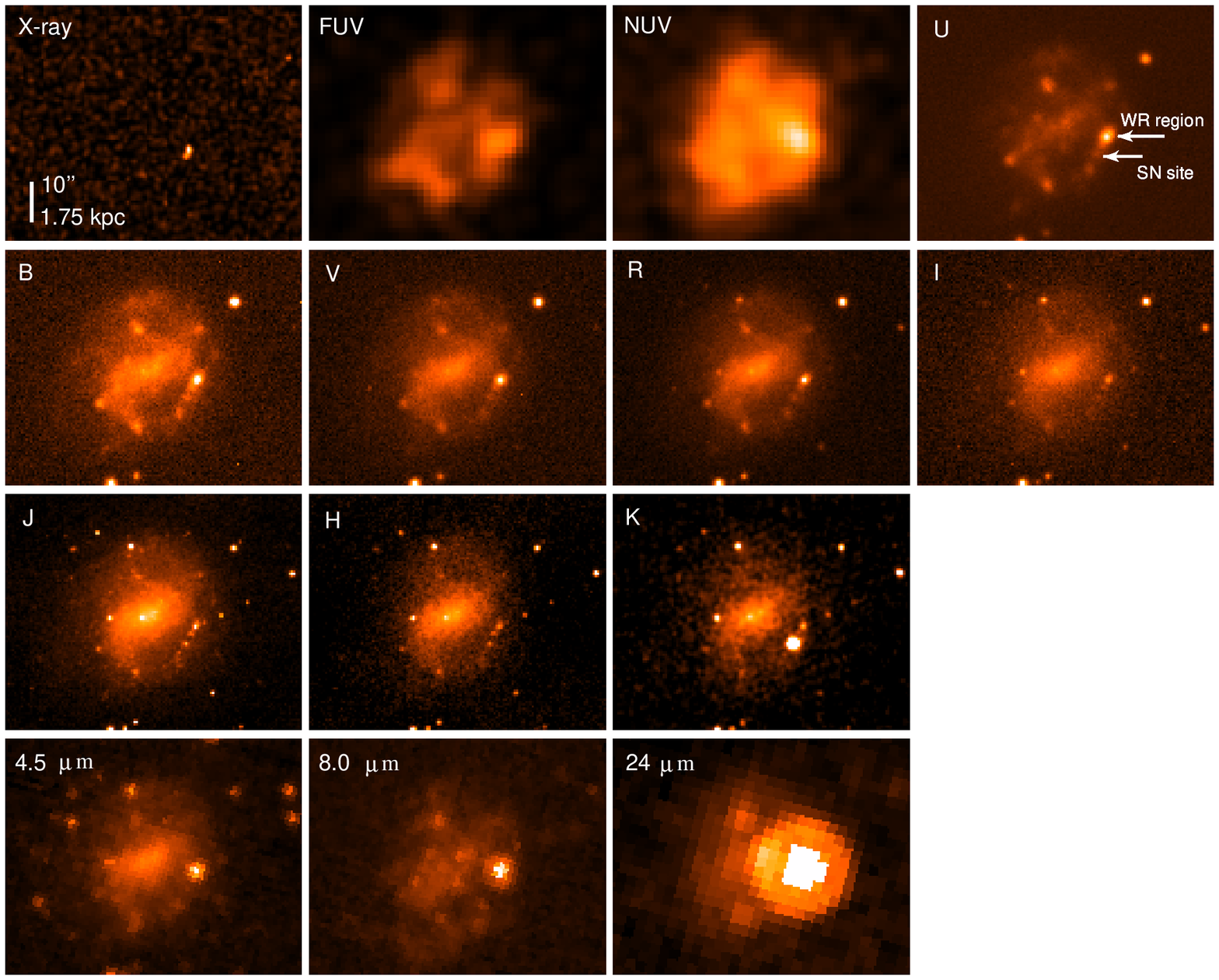}
\end{center}
\caption{Mosaic of images of \protect\object{ESO 184-G82}, the host galaxy of \protect\object{GRB 980425} / \protect\object{SN 1998bw}. North is up and east is to the  left. Images are $70''\times60''$ ($12\times10$ kpc at the redshift of $0.0085$). The scale is also shown on the first panel. From top left to bottom right: X-ray, far-UV, near-UV, $U$, $B$, $V$, $R$, $I$, $J$, $H$, $K$, $4.5$ \micron, $8.0$ \micron\, and $24\,\micron$. The arrows on the $U$-band image mark the SN site and the  WR region. The WR region is bright in the UV and mid-IR and faint in the near-IR (see Figure~\ref{fig:sed}) hinting at a very young  stellar population with overall small mass (compare with Table~\ref{tab:grasilres}). Note that the $K$-band image was obtained when the SN was still bright. The X-ray image reveals two compact sources $1.5''$ apart (overlapping at the image shown): the SN and an ultra-luminous X-ray source \citep{kouveliotou04}.}
\label{fig:image}
\end{figure*}

\newcommand{\GiveRef}[1]{\citetalias{#1}: \citet{#1}}
\defcitealias{watson04}{1}
\defcitealias{castroceron08}{3}
\defcitealias{sollerman05}{4}
\defcitealias{lefloch}{5}

\begin{table*}
\begin{center}
\caption{Results of the photometry of the GRB 980425 host and the WR region.}
\label{tab:sed}
\begin{footnotesize}
\begin{tabular}{l  cccccccccccccccc}
\hline\hline
Filter & X-ray & FUV & NUV & $U$ & $B$ & $V$ & $R$ & $I$ & $J$ & $H$ & $K$ & \multicolumn{3}{c}{{\it Spitzer}} & \multicolumn{2}{c}{ATCA} \\
$\lambda$ (\micron)& $6$ keV & 	0.1516&	0.2267&	0.36	&	0.428&	 0.553&	0.656&	0.767&	1.25&	1.64&	2.17&	4.5&		8.0	&	24	&	3 cm		&	6 cm \\
\hline
Host		& 3.6$\times10^{-7}$&	1.26&	1.54 	&	2.44	&	4.5	& 	5.56 	&	6.80	&	8.21	&	9.8	&	8.8	&	6.5	&	2.95	&	11.9 &	27.3	&	$<0.18$	&	0.42	\\
\multicolumn{1}{r}{Error}& 1.6$\times10^{-7}$&	0.13&	0.16	&	0.26	&	\dots	& 	\dots	&	\dots	&	\dots	&	0.4	&	0.5	&	0.4	&	0.10	&	 0.3	&	0.2	&	\dots		&	0.05	\\
\hline
WR				& \dots & 0.095		& 0.120	& 0.170	& 0.116	& 0.162	& 0.135	& 0.0690	& 0.078	& 0.071	& 0.044	&	0.22	&	1.815&	21	&	\dots	& \dots\\		
\multicolumn{1}{r}{Error}		 & \dots & 0.023		& 0.017	& 0.016	& 0.006	& 0.008	& 0.006	& 0.0034	& 0.013	& 0.012	& 0.009	&	\dots	&	\dots	&	\dots	&	\dots	& \dots\\
\multicolumn{1}{r}{(\%)}& \dots & 7.5	&	7.8	&	7.0	&	2.6	&	2.9	&	2.0	&	0.8	&	0.8	&	0.8	&	0.7	&	7.5	&	15.3	&	76.9 & \dots & \dots\\
\hline
Ref.		& \citetalias{watson04} &	2 	 	&	2,\citetalias{castroceron08}		&	2		& 2,\citetalias{sollerman05} & 2,\citetalias{sollerman05} & 2,\citetalias{sollerman05} & 2,\citetalias{sollerman05}&	2&	2&	2,\citetalias{castroceron08} & \citetalias{lefloch} & \citetalias{lefloch} & \citetalias{lefloch} &2 &2 	
\\
\hline
\end{tabular}
\end{footnotesize}
\tablecomments{Flux densities are given in mJy and are corrected for Galactic extinction assuming $E(V-B)=0.059$ \citep{schlegel98} and the extinction curve of \citet{cardelli89}. The row marked by \% shows the percentage contribution of the WR region to the total galaxy emission. The upper limit is $3\sigma$ and errors are $1\sigma$. References: \GiveRef{watson04}, 2: This work, \GiveRef{castroceron08}, \GiveRef{sollerman05},  \GiveRef{lefloch}.}
\end{center}
\end{table*}

We undertook deep radio observations of the host galaxy of GRB 980425 on 2007 August 18 using the Australia Telescope Compact Array (proposal no.~C1651, PI: Micha\l owski) using the hybrid H168 configuration, with antennas positioned on both east-west and north-south tracks, and baselines of 60--4500 m. Simultaneous observations were made at 6 cm (4.8 GHz) and 3 cm (8.64 GHz), with a bandwidth of 128 MHz at each frequency. A total of 10.5 hr of data were obtained. Calibrator source \object{PKS B1934-638} was utilized to set the absolute flux calibration of the array as well as to calibrate phases and gains. Data reduction and analysis was done using the MIRIAD package \citep{miriad}. Antenna \#1 was excluded from the analysis due to phase instabilities, thus reducing the number of possible baselines from 15 to 10. Calibrated visibilities were Fourier transformed using ``robust weighting'', which combines high signal-to-noise ratio with enhanced sidelobe suppression. The final synthesized beam sizes for 6 and 3 cm images were $76''\times38''$ and $37''\times21''$, respectively, with root-mean-square (rms) values of $46$ and $27\,\mu$Jy beam$^{-1}$. The host galaxy, \object{ESO 184-G82}, was only detected at 6 cm. This is only the third $>3\sigma$ radio detection of a GRB host, after those of \object{GRB 980703} and \object{GRB 000418} \citep{bergerkulkarni,berger}. Note that the radio observations of \object{GRB 000301C} and \object{GRB 000911} were also reported to be $>3\sigma$ detection, but after removal of the afterglow signal, the significance of the host detections drops below $3\sigma$. As \object{ESO 184-G82} slightly overlaps Galaxy A, reported by \citet{foley06}, $\sim70''$ south (see Figure~\ref{fig:radio}), its flux density was estimated by simultaneous fitting of two two-dimensional Gaussian functions to the data with their centroids, sizes, and orientations as free parameters. The lack of residuals left after the subtraction of these two Gaussians rules out a significant contamination of the Galaxy A to the measured flux of the host. \object{ESO 184-G82} was not detected at 3 cm down to a $3\sigma$ limiting flux of $0.18$ mJy.

We obtained $U$-band photometry on the Danish 1.5m Telescope on La Silla during the period 2007 May-June. In total 3.75 hr were spent on the target. The data were reduced in a standard manner using IRAF \citep{iraf1,iraf2}. 

We performed photometry on archival $JHK$ images from NTT/SofI \citep{patat01}, VLT/ISAAC \citep{sollerman02}  and  Two Micron All Sky Survey  \citep[2MASS][]{2mass}\footnote{2MASS XSC Final Release (Two Micron All Sky Survey Extended Source Catalog) released on 2003 March 25; \url{http://www.ipac.caltech.edu/2mass/}}, as well as $BVRI$ images from VLT/FORS1 \citep{sollerman05} and UV images from {\it GALEX} \citep{galex1,galex2}\footnote{{\it Galaxy Evolution Explorer}; \url{http://galex.stsci.edu/}}. 
The flux was measured in an aperture of $50''$  diameter for the whole galaxy and 2.4--3.6'' (depending on the seeing of the particular image) for the WR region. The results of our photometry and the fluxes obtained from the literature are presented  in Table~\ref{tab:sed} and a mosaic of images is shown in Figure~\ref{fig:image}.

Finally we analyzed the X-ray (2--10 keV) image from \citet{kouveliotou04}. It was, however, not used in the modeling since our SED templates do not cover this wavelength regime.

\section{SED Modeling}
\label{sec:model}

In order to model the SEDs of  \object{ESO 184-G82} and of the WR region we utilized the set of 35\,000 SED models from \citet{iglesias07} developed in GRASIL \citep{silva98}\footnote{\url{http://adlibitum.oat.ts.astro.it/silva/default.html}.} based on numerical calculations of the radiative transfer within a galaxy. They cover a broad range of galaxy properties from quiescent to starburst. We scaled all the SEDs to match the observational data and chose the one with the lowest $\chi^2$ to derive the galaxy characteristics. 

The radio parts of model SEDs were scaled down by an appropriate factor to account for the decreased efficiency of nonthermal radio emission of dwarf galaxies \citep[see Equation~(\ref{eq:sfrradio}) and discussion in Section \ref{sec:sfr} and in][]{bell03}. Namely a dwarf galaxy has a lower radio flux than it would result from scaling down the high-luminosity SED template and the GRASIL model does not take into account this effect. From the  SFR--radio flux relation of \citet[][see equation~(\ref{eq:sfrradio}) below]{bell03} we inferred that the radio part of the SED template corresponding to \object{ESO 184-G82} should be $\approx3.5$ times lower than in the original template.
Anyway, even such corrected templates  overpredict the value of radio data points so we excluded them from the fitting (see Section \ref{sec:radio} for a discussion).
 
\begin{figure*}[p]
\begin{center}
\plotone{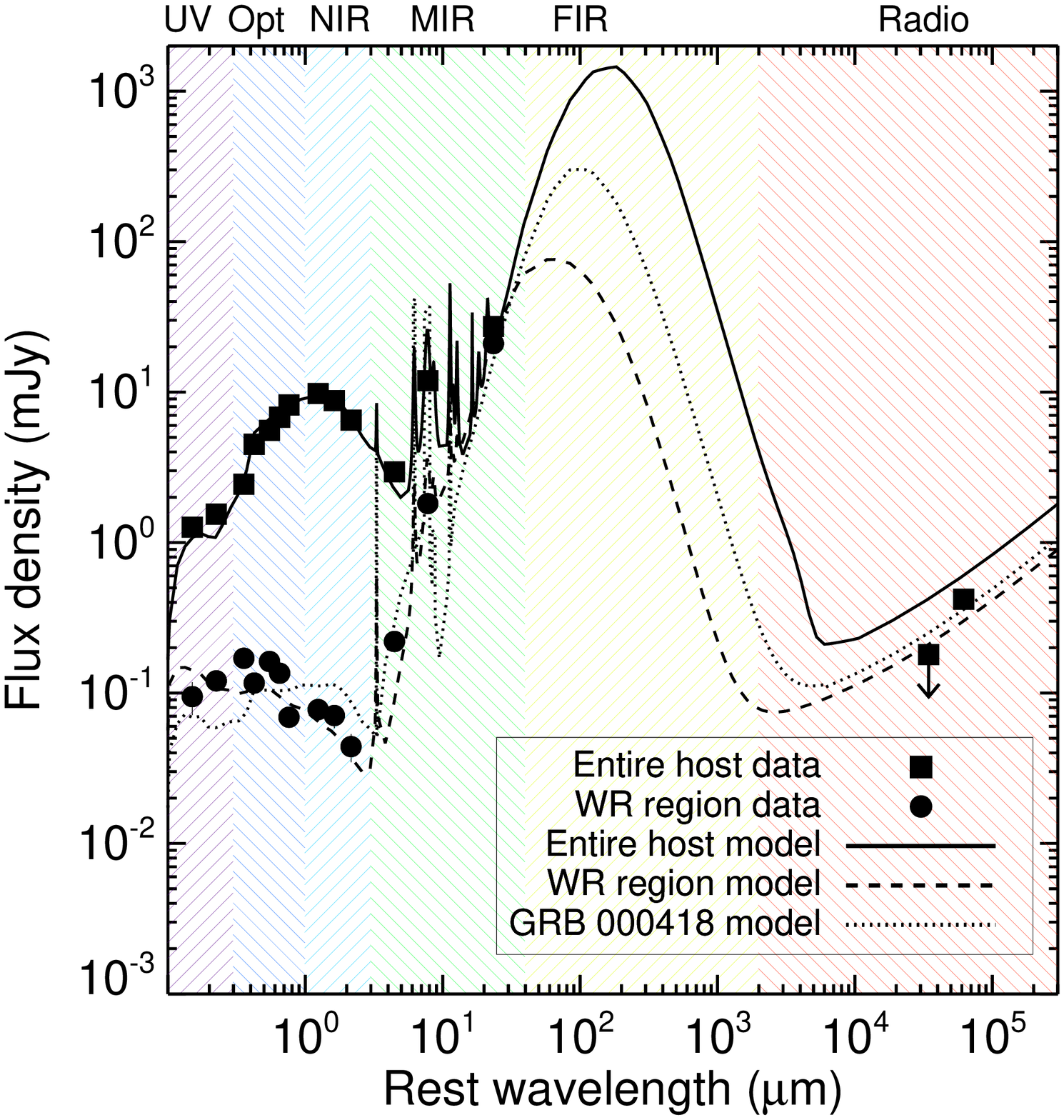}
\end{center}
\caption{Spectral energy distribution of \protect\object{ESO 184-G82}, the host of \protect\object{GRB 980425} / \protect\object{SN 1998bw}, compared to the model  corresponding to the host of GRB 000418 \citep{michalowski08}. {\it Solid line}: spiral galaxy model (entire host). {\it Dashed line}: young starburst model (WR region). Both models have been calculated using GRASIL \citep{silva98} by \citet{iglesias07}.  {\it Dotted line}: model based on the host of \protect\object{GRB 000418} from \citet{michalowski08} (slightly modified; see Section \ref {sec:wr}).
{\it Squares} and {\it circles}: detections of the host galaxy and the WR region, respectively, with errors, in most cases, smaller than the size of the symbols. {\it Arrow}: $3\sigma$ upper limit (values marked at the base).
The hashed columns mark the wavelength ranges corresponding
to the UV, optical, near-IR, mid-IR, far-IR, and radio domains. 
For a discussion of the discrepancy between the data and models at radio wavelengths see Section \ref{sec:radio}.}
\label{fig:sed}
\end{figure*}

 The best fits\footnote{The SED fits can be downloaded from\\ \protect\url{http://archive.dark-cosmology.dk}} 
are shown in Figure~\ref{fig:sed} and the resulting properties of the galaxy are listed in Table~\ref{tab:grasilres} \citep[see][ and Sections \ref{sec:sfr} and \ref{sec:dust} for details on how these were derived from the SEDs]{michalowski08}.

\section{Results}
\label{sec:results}

\begin{table*}
\caption{Properties of the host galaxy ESO 184-G82 and its WR region derived from the SED modeling.}
\begin{center}
\begin{tiny}
\begin{tabular}{lcccccccccccc}
\hline\hline
 & Age 	& $L_{\rm IR}$		& SFR$_{\rm SED}$ & SFR$_{\rm UV}$ & SFR$_{\rm IR}$ & SFR$_{\rm radio}$ & $M_*$ & $M_{\rm burst}$ & $M_{\rm dust}$ & $T_d$ & $A_{\rm 1 \micron}^{\rm MC}$ & $A_V^{\rm av}$\\ 
Region & (Gyr)	& ($10^{9}L_\odot$)	& ($M_\odot$ yr$^{-1}$) & ($M_\odot$ yr$^{-1}$) & ($M_\odot$ yr$^{-1}$) & ($M_\odot$ yr$^{-1}$) & ($10^{9}\,M_\odot$) & ($10^{6}\,M_\odot$) & ($10^{5}\,M_\odot$) & (K) & (mag) & (mag) \\
 (1) &  (2) & (3) & (4) & (5) & (6) & (7) & (8) & (9) & (10) & (11) & (12) & (13)\\ 
 \hline
Host & $12.00^{+0.00}_{-4.00}$ & $2.64^{+0.13}_{-0.31}$ & $0.38^{+0.05}_{-0.05}$ & $0.25^{+0.04}_{-0.06}$ & $0.45^{+0.02}_{-0.05}$ & $0.23\pm0.04$ & $1.6^{+0.25}_{-0.10}$ & \dots & $28^{+72}_{-14}$ & $38$ & $0.34$ & $0.07$\\
WR &  $3.56^{+8.44}_{-1.98}$ & $0.36^{+0.28}_{-0.26}$ & $0.10^{+0.05}_{-0.05}$ & $0.03^{+0.01}_{-0.02}$ & $0.06^{+0.05}_{-0.04}$ & \dots & $0.0057^{+0.0063}_{-0.0025}$ & $4.5^{+2.4}_{-2.4}$ & $0.05^{+0.14}_{-0.03}$ & $92$ & $0.78$ & $0.04$\\
\multicolumn{1}{l}{(\%)} & \dots & $14$ & $26$ & $12$ & $13$ & \dots & $0.4$ & \dots & 0.2 & \dots & \dots & \dots \\
\hline
\end{tabular}
\label{tab:grasilres}
\end{tiny}
\tablecomments{Column~(1):~the entire host galaxy / only WR region / percentage contribution of the WR region to the cumulative properties of the galaxy. Column~(2):~age, defined as the time since the beginning of its evolution. Column~(3):~total $8-1000$ \micron\, infrared luminosity. Column~(4):~total star formation rate (SFR) for $0.15-120\,M_\odot$ stars averaged over the last 50 Myr derived from the SED model. Column~(5):~SFR from UV emission \citep[Table~\ref{tab:sed}, using][]{kennicutt}. Column~(6):~SFR from IR emission  \citep[Column 3, using][]{kennicutt}. Column~(7):~SFR from radio emission  \citep[Table~\ref{tab:sed}, using][]{bell03}. Column~(8):~stellar mass.
Column~(9):~mass of gas converted into stars during starburst. Column~(10):~dust mass. Column~(11):~dust temperature. Column~(12):~extinction in molecular cloud (MC) at 1 \micron\, measured from its center. Column~(13):~average extinction of stars outside MCs at 0.55 \micron.}
\end{center}
\end{table*}

\subsection{Stellar Masses}
\label{sec:mstar}

The broadband SED of the host of \object{GRB 980425} is consistent with that of a galaxy with an old stellar population (the time since the beginning of the galaxy evolution is equal to $12$ Gyr, see Column 2 of Table~\ref{tab:grasilres}) built up quiescently without any starburst episode \citep[consistent with the conclusion of][]{sollerman05} at a rate comparable to the present value. 
The age estimate is however uncertain due to degeneracy between age and dust extinction as well as metallicity, namely that if one increases the assumed metallicity or decreases the extinction then the resulting age will increase.
The derived stellar mass agrees with previous estimates \citep{castroceron08,savaglio09}.

On the other hand, the comparison of Columns 8 and 9 of Table~\ref{tab:grasilres} reveals that the stellar mass of the WR region is dominated by a starburst episode, so that it has built up a negligible fraction of the stellar mass before the starburst. According to our SED model, this starburst is still ongoing and started $50$ Myr ago. Interestingly this is the starburst age predicted for GRB hosts by \citet{lapi08} based on the argument that for older starbursts the metallicity becomes too high to produce a GRB.

\subsection{Star Formation Rates}
\label{sec:sfr}

The SFR of the entire galaxy, as well as that of the WR region, was calculated from UV and infrared (IR) fluxes (Table~\ref{tab:sed}) using the conversions of \citet{kennicutt}. 
The radio SFR ($M_\odot$ yr$^{-1}$) was  calculated from the radio luminosity $L_{\rm 1.4\, GHz}$ (erg s$^{-1}$ Hz$^{-1}$) using the method proposed by \citet{bell03}:
\begin{equation}
{\rm SFR}=\left\{
\begin{array}{ll}
5.52\times10^{-29} L_{\rm 1.4\, GHz} &  L_{\rm 1.4\, GHz} > L_c\\
\displaystyle\frac{5.52\times10^{-29} L_{\rm 1.4\, GHz}}{0.1+0.9(L_{\rm 1.4\, GHz}/L_c)^{0.3}} &  L_{\rm 1.4\, GHz} < L_c,\\
\end{array}
\right.
\label{eq:sfrradio}
\end{equation}
where $L_c=6.4\times10^{28}$ erg s$^{-1}$ Hz$^{-1}$ is a critical luminosity (see below). This relation was derived based on a sample of 249 galaxies spanning a wide range in luminosities including normal and intensely star-forming galaxies, starbursts, ultraluminous IR galaxies and blue compact dwarfs. The luminosity at the rest frequency of $1.4$~GHz, $L_{\rm 1.4\, GHz}$~(erg s$^{-1}$ Hz$^{-1}$), of a galaxy at redshift $z$ and luminosity distance $D_L$~(cm), can be calculated from the flux density $F_\nu$~(Jy) at the observed radio frequency $\nu_{\rm obs}$~(GHz) assuming the radio spectral slope $\alpha=-0.75$ \citep{yun}:
\begin{equation}
L_{\rm 1.4\, GHz}=\frac{4\pi\times 10^{-23} D_L^2 F_\nu}{1+z} \times \left[\frac{\nu_{\rm obs}(1+z)}{1.4}\right]^{-\alpha}.
\label{eq:lumradio}
\end{equation}
This relation (Equation~(\ref{eq:sfrradio})) diverges significantly from the usual methods \citep{condon,yun} for low-luminosity galaxies, because the nonthermal radio emission is not effective in such galaxies  and the relation between SFR and radio luminosity becomes nonlinear below $L_c$ (SFR$\mbox{}\lesssim3\, M_\odot$ yr$^{-1}$). This effect is likely caused either by the fact that cosmic-ray electrons responsible for the radio emission escape from galaxies of small sizes \citep{bell03} or that the ordered magnetic field in dwarf galaxies is weaker and therefore magnetic field due to SNe (responsible for acceleration of  electrons) is less efficient because it results from contraction and amplification of the global field.

The SFR derived from SED modeling (Column 4 of Table~\ref{tab:grasilres}) agrees (within a factor of 2) with the estimates derived from UV, IR, and radio for the entire galaxy, suggesting little extinction  (see also Section \ref{sec:dust}). All the estimates are also consistent with the X-ray SFR upper limit of $2.8\,M_\odot$ yr$^{-1}$ derived by \citet{watson04}. 

As noted by \citet{lefloch} the contribution of the WR region to the galaxy luminosity at $24$ \micron\, is $\sim75$\% (see Table~\ref{tab:sed} and Figure~\ref{fig:sed}). However, according to our SED fit, it only emits 15\% of the total IR luminosity (it would require high-resolution far-IR or submillimeter imaging to confirm this result). Under the assumption that the total IR luminosity is proportional to the SFR \citep{kennicutt}, this is consistent within a factor of 2 with the finding of \citet{sollerman05} and \citet{christensen08} that the WR region harbors about one-third of the host star formation (as also suggested by the SFRs derived directly from SED fits, see Column 4 of Table~\ref{tab:grasilres}).

\subsection{Dust Properties}
\label{sec:dust}

We derived the dust temperature by fitting a graybody curve to the model SED near the dust peak \citep[as in][]{michalowski08}. The dust in the WR region is much hotter than the average over the entire galaxy (see Column 11 of Table~\ref{tab:grasilres} and note on Figure~\ref{fig:sed} that the SED of the WR region peaks at shorter wavelengths than that of the entire galaxy). This hints at a very intense starburst episode and a strong radiation field, consistent with the discussion in Section \ref{sec:mstar}. High dust temperatures are not uncommon for GRB hosts. They were found  for higher-redshift ($z=0.9$--$1.5$) GRBs with similar conclusions about their origin \citep{michalowski08}. Moreover, \citet{bloom03b} and \citet{djorgovski01} noted that high flux ratios between the [\ion{Ne}{3}] and [\ion{O}{2}] lines in GRB hosts suggest the presence of very hot \ion{H}{2}  regions. 
 
The total dust mass,  $M_d$,  was estimated using the method of \citet{taylor} based on the formalism developed by \citet{hildebrand}:
\begin{equation}
M_d=\frac{F_\nu D_L^2}{(1+z)\kappa(\nu_r)B(\nu_r,T)},
\label{eq:mdgeneral}
\end{equation}
where $F_\nu$ is the flux density (either observed or
interpolated from an SED model)  at the rest frequency dominated by dust thermal emission ($200\,\mbox{GHz}\lesssim\nu_r\lesssim 5000\,\mbox{GHz}$),  $B(\nu_r,T)$ is the Planck function, $T$ is dust temperature, $\kappa$ is the mass absorption coefficient
\begin{equation}
\kappa(\nu_r)=0.067\left[\frac{\nu_r\,(\mbox{GHz})}{250}\right]^\beta \quad \mbox{m}^2\,\mbox{kg}^{-1}
\label{eq:kappa}
\end{equation}
 and $\beta$  is the emissivity index.
Equations (\ref{eq:mdgeneral}) and (\ref{eq:kappa}) can be combined into the following easy-to-use formula where the resulting dust mass $M_d$ ($M_\odot$) is computed from the quantities in the following units: $F_\nu$ (Jy), $D_L$ (cm), $\nu_{\rm obs}$ (GHz), and $T$ (K):
\begin{equation}
M_d=3.24\times10^{-44}\frac{F_\nu D_L^2 \left\{\exp\left[ \frac{0.048\nu_{\rm obs}(1+z)}{T} \right] -1 \right\} }    {(1+z)\left[\frac{\nu_{\rm obs}(1+z)}{250} \right]^{\beta+3}}.
\label{eq:md}
\end{equation}

We estimated the flux at $450$ \micron\, from the SED models. The results of dust masses are given in Column 10 of Table~\ref{tab:grasilres} assuming $\beta=1.3$. There exists a degeneracy between the value of this parameter and resulting dust mass in a way that more dust is expected if lower $\beta$ is assumed. The uncertainties quoted in Table~\ref{tab:grasilres} are large, because we allowed a broad range of $\beta$  \citep[$0-2$;][]{yun}.

\citet{hatsukade07} derived an upper limit on the molecular mass of the host of \object{GRB 980425} $M_{{\rm H}_2}<3\times10^8\,M_\odot$. Therefore from our dust mass estimate  we derive a molecular gas-to-dust ratio $M_{{\rm H}_2}/M_d<107$. This value is lower than the molecular gas-to-dust ratio for the Milky Way  \citep[$\sim140-400$;][]{sodroski97,draine07} and other spirals  \citep[$\sim1000 \pm 500$;][]{devereux90,stevens05}, but consistent with the values for high-redshift submillimeter galaxies \citep[$54_{-11}^{+14}$;][]{kovacs06}, and for the nuclear regions of local luminous IR galaxies (LIRGs), ultraluminous IR galaxies (ULIRGs) \citep[$120\pm28$;][]{wilson08} and of local, far-IR-selected galaxies \citep[$\sim50$;][]{seaquist04}. It indicates that the host of \object{GRB 980425} harbors a relatively large amount of dust, or that its gas reservoir is significantly depleted.
However this conclusion is based on an uncertain dust mass estimate, so should be checked with deep submillimeter observations. 

Our SED fits are consistent with negligible extinction for both the entire galaxy and the WR region (Column 13 of Table~\ref{tab:grasilres}). Very low reddening for the entire galaxy was also found by \citet{patat01} and \citet{sollerman05} from the width of the \ion{Na}{1} D doublet and SED fitting, respectively. On the other hand, using the Balmer decrement, \citet{savaglio09} and \citet{christensen08} derived $A_{\rm V}=1.73$ and $0.93$, respectively, for the entire galaxy, whereas \citet{hammer06} and \citet{christensen08} obtained  $A_{\rm V}=1.51$ and $0.53$, respectively, for the WR region. However, extinction derived from emission lines of the \ion{H}{2} regions is usually higher than from the SED modeling \citep{savaglio09}.

\section{Discussion}
\label{sec:discussion}

\subsection{The Host Galaxy}

From the SED modeling it is apparent that \object{ESO 184-G82}, the host galaxy of \object{GRB 980425} / \object{SN 1998bw},  is a normal dwarf star-forming spiral. None of its properties (Table~\ref{tab:grasilres}) are exceptionally high or low. In particular its mass, SFR, and size are broadly consistent with the range obtained for a sample of local dwarf galaxies \citep[Fiigures~5 and 17 of][in this respect \object{ESO 184-G82} is very similar to the \object{Large Magellanic Cloud}]{woo08} and for a sample of local blue compact galaxies \citep[Figure~2 of][]{sollerman05}. 

Its specific SFR ($\phi\equiv{\rm SFR}/M_*=0.23$~Gyr$^{-1}$)  
 is consistent with the range of $\phi$ found for other GRB hosts by \citet{castroceron08} based on UV \citep[but lower than for a subsample detected in mid-IR;][]{castroceron06}. However its $\phi$ is higher than for the majority of  nearby spiral galaxies hosting SNe  \citep[see Figure~8 of][]{thone08b}.
High $\phi$  for other GRB hosts were also reported by, e.g., \citet{christensen04} and predicted theoretically by \citet{courty04,courty07} and \citet{lapi08}. This is in agreement with the finding of \citet{iglesias06,iglesias07} and \citet{zheng07} that low-mass galaxies  in general have high $\phi$.

As stated in Section \ref{sec:mstar}, the SED of \object{ESO 184-G82} is consistent with being of a nonstarburst nature. This is also supported by its  stellar building time ($T_{\rm SFR}\equiv \phi^{-1}=4$ Gyr) being not much less than the Hubble time and low SFR per unit area equal to $ 0.004\,M_\odot$ yr$^{-1}$ kpc$^{-2}$ \citep[see the relevant discussion in][]{heckman05}. 

\object{ESO 184-G82} is the only GRB host with a clear $\sim1.6\,\micron$ bump  in the SED \citep[compare Figure~\ref{fig:sed} with Figure~4 of][]{savaglio09}. According to \citet{sawicki02} this feature starts to be apparent for a galaxy older than $~100$ Myr (see his Figure~1). The preference of not having the bump for other GRB hosts likely indicates that on average they are very young galaxies, although we stress that in many cases the optical and near-IR data presented by \citet{savaglio09} do not cover the wavelengths into where the bump is redshifted.

\subsection{Radio Detection}
\label{sec:radio}

The SED model presented in Figure~\ref{fig:sed} (solid line) overpredicts the radio fluxes by a factor of $1.5$ ($>2.3$) in the $6$ ($3$) cm band. We suggest that this may result from the following effect. Radio wavelengths probe current star formation activity \citep[$\lesssim10^8$ yr;][]{condon,cannon04}, unlike UV \citep{kennicutt,christensen04} and IR \citep{calzetti07}, at which even older galaxies can be luminous. Therefore it seems likely that only a limited part of the galaxy is younger than $10^8$ yr, so the galaxy is fainter in radio than its UV and IR fluxes would imply. This is supported by \citet{sollerman05} who noticed that the colors of the \object{GRB 980425} host are consistent with a constant SFR over 5--7 Gyr without any starburst episode. Therefore if we assume that the IR probes the total SFR, then the radio data point would be a factor of $\sim2$ ($\approx\mbox{SFR}_{\rm IR}/\mbox{SFR}_{\rm radio}$) higher if the radio were also sensitive to star formation older than $10^8$ yr. 

We calculated the radio spectral index $\alpha$ defined as $F_\nu\propto\nu^\alpha$, so $\alpha_{\nu_1}^{\nu_2}= \log [F_\nu(\nu_2)/F_\nu(\nu_1)] / \log (\nu_2/\nu_1)$. The radio SED of \object{ESO 184-G82} (see Table~\ref{tab:sed}) is very steep with $\alpha^{8.64}_{4.8}<-1.44$.
This is consistent with the steepest slopes in the sample of ULIRGs discussed by \citet{clemens08} and interpreted as an indication of spectral aging of relativistic electrons (the lifetime of high-energy electrons emitting high-frequency radiation is shorter than for low-energy electrons). The same conclusion is drawn by  \citet{hirashitahunt06} who predicted a steepening of the radio slope  $\sim10$ Myr after a starburst when synchrotron radiation starts to dominate over  free-free emission from \ion{H}{2} regions  \citep[see also][]{bressan02,cannon04}. In summary, such a steep radio slope indicates that the bulk of star formation activity in the host of \object{GRB 980425} is not recent.

As mentioned in Section \ref{sec:sfr} the radio SFR for dwarf galaxies can underpredict the true value if derived using usual methods.  Since GRB hosts are in general subluminous at all wavelengths \citep{hoggfruchter99,hanlon,hjorth00,hjorth02,vreeswijk01,fynbo02,fynbo03,fynbo06,berger,lefloch03,lefloch,christensen04,courty04,tanvir,tanvir08,jakobsson05,sollerman05,sollerman06,fruchter06,priddey06,chary07,ovaldsen07,thone07b,wiersema07,castroceron08,savaglio09} we suggest that the \citet{bell03} relation (Equation~(\ref{eq:sfrradio})) should be used to calculate their radio SFRs.  Indeed, in the case of the host of \object{GRB 980425}, one would get a value of $0.068\, M_\odot$ yr$^{-1}$ using the method of \citet{yun}, a value much smaller than the UV SFR. The radio luminosity is supposed to trace both unobscured and obscured SFRs (because radio is not affected by dust), so such a low value is clearly an underestimation of the true SFR. The relation of \citet{bell03} is however not necessary (but gives reasonable results) for the high-luminosity subsample of GRB hosts where usual methods result  in radio SFRs consistent with other diagnostics \citep[see Table~1 of][]{michalowski06}.

\subsection{WR Region}
\label{sec:wr}

The WR region emits $7$\% of the host's UV flux. Its contribution falls to below $1$\% in the near-IR and rises steeply to $75$\% in the mid-IR. As mentioned in Sections \ref{sec:mstar} and \ref{sec:dust}, an intense starburst episode together with low stellar mass provide a consistent explanation of the shape of the SED. Indeed our SED fit suggests that the WR region harbors as much as 12--26\% of the total star formation activity, but its contribution to the galaxy stellar and dust masses is negligible (see Columns 8 and 10 of Table~\ref{tab:grasilres}).  

The $\phi$ of the WR region is $22$ Gyr$^{-1}$.  
 High $\phi$ in the immediate environment of GRBs was also found by \citet[][see their Figure~4; the spatial resolution was $3$ kpc in this case]{thone08} and is consistent with the findings of  \citet{fruchter06}. We stress that we do not claim here that GRB 980425 is physically connected with the WR region, just that it occurred in the most intense star-forming part of the galaxy \citep[note in Figure~\ref{fig:image} that the Southern spiral arm is the only part of the galaxy where X-ray point sources are found, indicative of intense star formation;][]{kouveliotou04}. Because of the proximity of the SN region to the WR region, it is  very likely that their star formation was triggered by the same  mechanism and therefore that the nature of their star formation is similar.
 
 The starburst nature of the WR region is confirmed by its stellar building time ($T_{\rm SFR}=57$ Myr) being  much less than the Hubble time, and its very high SFR per unit area equal to $6\,M_\odot$ yr$^{-1}$ kpc$^{-2}$ \citep{heckman05}.

It is worth noting that the SED of the WR region is qualitatively similar to the SEDs presented by \citet{michalowski08} for submillimeter/radio-bright GRB hosts: blue in the optical, luminous in the mid-IR, and indicating hot dust content. The similarities are highlighted  in Figure~\ref{fig:sed} where the WR region model (dashed line) and the model corresponding to \object{GRB 000418} (dotted line)   are compared. The agreement is striking, but note that in order to suppress the very high IR luminosity of the host of \object{GRB 000418}, we needed to modify the SED model presented by \citet{michalowski08} by changing the escape parameter from $50$ to $10$ Myr \citep[the time after which stars begin to escape from molecular clouds; see][for a discussion of this parameter]{panuzzo07}.

The WR region was also found to be similar to high-$z$ GRB hosts with respect to emission line ratios (indicative of age and metallicity), unlike the entire host galaxy \object{ESO 184-G82}, which appears to be older than other GRB hosts \citep{christensen08}.

The picture that emerges from these findings is that the $\sim$1--3 kpc scale environment of a GRB represents the youngest and most intensely star-forming region of a host galaxy, harboring the hottest dust. If present at high redshifts such regions may dominate the emission (and therefore, derived properties) of distant GRB hosts. 

\section{Conclusions}
\label{sec:conclusion}

In this paper we have presented the UV-to-radio SED fitting of the host galaxy of \object{GRB 980425} / \object{SN 1998bw} and of the WR region close to the SN position.

The host galaxy of \object{GRB 980425} is a normal dwarf spiral galaxy with somewhat elevated star formation activity compared to other spirals (though it is not necessary to invoke any starburst episode to explain its SED). The steep radio slope and the presence of the $\sim1.6\,\micron$ bump in the SED indicate the existence of an old stellar population.  Its low radio luminosity can be explained by the suppression of synchrotron emission in dwarf galaxies and the fact that radio is only sensitive to recent star formation. 
 
The emission of the WR region close to the GRB position is dominated by an ongoing starburst episode, during which almost all of its stars were formed. It contributes significantly  to the star formation of the entire galaxy. In many aspects the WR region is similar to high-redshift GRB hosts: it is a blue, young region of intense star formation containing hot dust. The presence of the GRB close to this region indicates that GRBs appear to be  associated with regions of high specific SFR and high dust temperatures.

\acknowledgments
We thank Joanna Baradziej, Eelco van Kampen, Robin Wark, Mark Wieringa for discussion and comments; our referee Sandra Savaglio for help in improving this paper; Jorge Iglesias-P\'{a}ramo for kindly providing his SED templates; Naomi McClure-Griffiths for help with ATCA observations; Paul M.~Vreeswijk, Jesper Sollerman, Ferdinando Patat, and Chris Lidman for providing the reduced optical and near-infrared images; Andreas O.~Jaunsen for help on the data reduction; and Laura Silva for making the GRASIL code available.

The Dark Cosmology Centre is funded by the Danish National Research Foundation. M.~J.~M.~would like to acknowledge support from The Faculty of Science, University of Copenhagen.
J.~M.~C.~C.~gratefully acknowledges support from the Instrumentcenter for Dansk Astrofysik and the Niels Bohr Institutet's International PhD School of Excellence.
The Australia Telescope Compact Array is part of the Australia Telescope which is funded by
the Commonwealth of Australia for operation as a National Facility managed by CSIRO. The authors acknowledge the data analysis facilities provided by the Starlink Project which is run by CCLRC on behalf of PPARC. This research has made use of the NASA's Astrophysics Data System; the GHostS database ({\tt http://www.grbhosts.org/}), which is partly funded by {\it Spitzer}/NASA grant RSA Agreement No.~1287913; the Gamma-Ray Burst Afterglows site ({\tt http://www.mpe.mpg.de/$\sim$jcg/grb.html}), which is maintained by Jochen Greiner; IRAF, distributed by the National Optical Astronomy Observatories, which are operated by the Association of Universities for Research in Astronomy, Inc., under cooperative agreement with the National Science Foundation; the NASA/IPAC Extragalactic Database (NED) which is operated by the Jet Propulsion Laboratory, California Institute of Technology, under contract with the National Aeronautics and Space Administration; and SAOImage DS9, developed by Smithsonian Astrophysical Observatory. {\em Galaxy Evolution Explorer} ({\em GALEX} ) is a NASA Small Explorer, launched in 2003 April. We gratefully acknowledge NASA's support for construction, operation, and science analysis for the {\em GALEX} mission, developed in cooperation with the Centre National d'Etudes Spatiales of France and the Korean Ministry of Science and Technology.

\end{document}